\documentclass[
  12pt,
  a4paper,
  final
]{iopart}

\usepackage{iopams}  
\usepackage{graphicx}
\usepackage[breaklinks=true,colorlinks=true,linkcolor=blue,urlcolor=blue,citecolor=blue]{hyperref}

\expandafter\let\csname equation*\endcsname\relax
\expandafter\let\csname endequation*\endcsname\relax
\usepackage{amsmath}
\usepackage{amssymb}
\usepackage{color}
\usepackage{graphicx}

\newcommand{\sect}[1]{Sect.~\ref{#1}}
\newcommand{\fig}[1]{Fig.~\ref{#1}}

\newcommand{\eq}[1]{Eq.~(\ref{#1})}

\renewcommand{\epsilon}[0]{\varepsilon}

\begin{document}

\title[A general purpose tool for the construction of atomic interaction models]{\textsc{atomicrex} -- A general purpose tool for the construction of atomic interaction models}

\newcommand{\tudarmstadt}{
  Technische Universit\"at Darmstadt,
  Institut f\"ur Materialwissenschaften,
  64287 Darmstadt, Germany
}
\newcommand{\chalmers}{
  Chalmers University of Technology,
  Department of Physics,
  S-412 96 Gothenburg, Sweden
}

\author{Alexander Stukowski}
\ead{stukowski@mm.tu-darmstadt.de}
\address{\tudarmstadt}

\author{Erik Fransson}
\address{\chalmers}

\author{Markus Mock}
\address{\tudarmstadt}

\author{Paul Erhart}
\ead{erhart@chalmers.se}
\address{\chalmers}

\begin{abstract}
We introduce \textsc{atomicrex}, an open-source code for constructing interatomic potentials as well as more general types of atomic-scale models. Such effective models are  required to simulate extended materials structures comprising many thousands of atoms or more, because electronic structure methods become computationally too expensive at this scale. \textsc{atomicrex} covers a wide range of interatomic potential types and fulfills many needs in atomistic model development. As inputs, it supports experimental property values as well as \textit{ab initio} energies and forces, to which models can be fitted using various optimization algorithms. The open architecture of \textsc{atomicrex} allows it to be used in custom model development scenarios beyond classical interatomic potentials while thanks to its Python interface it can be readily integrated e.g., with electronic structure calculations or machine learning algorithms.
\end{abstract}

\maketitle

\section{Introduction}

Atomic-scale modeling plays an important role in analyzing, understanding, and predicting materials behavior. Many phenomena and processes of interest in this context involve length  scales that require simulations of tens of thousands to several millions of atoms over time scales that span nanoseconds or more. The application of electronic structure methods for this purpose --commonly referred to as \textit{ab initio} models-- is not only prohibitively expensive from a computational standpoint but many times even unnecessary as the relevant interactions do not require an explicit description of the electronic structure. This is the realm of semi-empirical methods, most notably interatomic potentials, for which the computation of interatomic forces is not only several orders of magnitude faster than with popular electronic structure methods such as density functional theory \cite{Jon15} (DFT) but also exhibits a much more favorable scaling with system size.

In contrast to e.g., DFT, where generic exchange-correlation functionals are available that can be applied to materials of almost arbitrary composition, interatomic potentials are typically constructed for specific elements and compounds, and often with particular applications in mind. Since the development of interatomic potentials is frequently a tedious and time consuming process, it requires tools that are both efficient and flexible. While various potential development tools have been developed for internal use by research groups, relatively few have been made widely available including e.g., \textsc{potfit} \cite{BroGah07}, \textsc{GARFfield} \cite{JaramilloBotero2014}, \textsc{MEAMfit} \cite{Duff2015}, the ``EAM Alloy Potential Generator'' \cite{WarAgrFlo12}, and the \text{aenet} package for artificial neural network (ANN) potentials \cite{ArtUrb16}. Several of these codes target specific potential types and/or functional forms; also they can be difficult to extend and/or integrate with other processing pipelines, in particular the popular Python scripting language.

The complexity of systems targeted by atomic scale modeling continues to increase while the possibilities for machine learning and systematic model construction multiply. Thus there is a need for efficient and flexible potential construction tools that are extensible and can be integrated in more complex model development setups. In response to this need we have developed the \textsc{atomicrex} code, the main features of which are
\begin{itemize}
\item
  support for a variety of interatomic potential forms,
\item
  the possibility for the user to define arbitrary functional forms via a built-in math expression parser,
\item
  very high computational efficiency enabling large training and validation sets,
\item
  a range of predefined properties that can be combined to generate more complex properties, in particular energy differences, defect energies, etc., which can be included in both training and validation stages,
\item
  an interface to the popular Python programming language, which enables interfacing with various third-party libraries, and
\item
  an object-oriented code framework that readily allows addition of new potential models, structures, and properties.
\end{itemize}
The models created using \textsc{atomirex} can be used for e.g., in simulation codes such as \textsc{lammps} \cite{Pli95} or \textsc{atomistica} \cite{atomistica} but can also be made available to the scientific community e.g., via the Knowledgebase for Interatomic Potentials \cite{kim, TadEllSet11}.

In the following we first provide a rough outline of the \textsc{atomicrex} workflow from a user perspective (\sect{sect:work-flow}) and introduce some key concepts (\sect{sect:key-concepts}). We continue by summarizing the optimization algorithms (\sect{sect:optimization-algorithms}) and potential formats (\sect{sect:potentials}) that are currently supported as well as the specification of atomic structures (\sect{sect:structures}) and user-defined properties (\sect{sect:derived-properties}).
We close with an outlook indicating directions of possible future developments (\sect{sect:conclusions-and-outlook}).

\textsc{atomicrex} is available under the GNU General Public License and is hosted in a public Git repository on GitLab, where its source code is available for download and which can be accessed via \url{http://atomicrex.org}. Most of the code base is written in C++, with Python bindings enabling integration with third-party libraries and custom model fitting schemes. An extensive user guide that contains a comprehensive description of features, input file parameters, and the Python interface along with various examples is available online. A separate documentation of the C++ and Python application programming interfaces (APIs) are available as well.

\section{Workflow}
\label{sect:work-flow}

\begin{figure*}
    \centering
    \includegraphics[width=0.98\columnwidth]{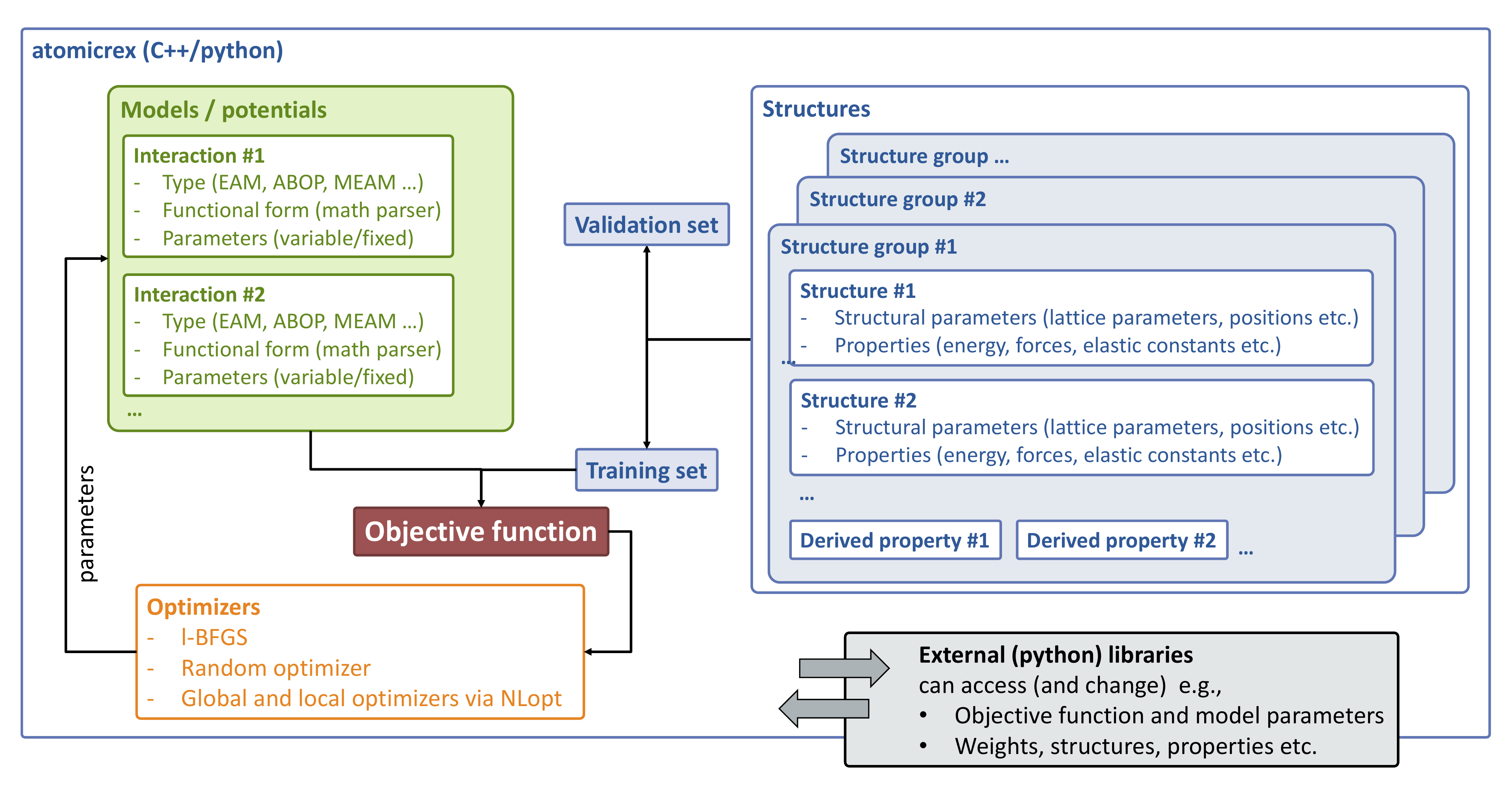}
    \caption{
        Schematic overview of the different objects handled by \textsc{atomicrex} and their connections.
    }
    \label{fig:overview}
\end{figure*}

From a technical viewpoint, \textsc{atomicrex} processes a user-supplied input file in extended markup language (XML) format that describes the job to be performed. Here, we will not discuss the specific format of this file or any of the parameters it contains, as a very extensive user guide including various examples is available online. In the following, we rather emphasize key steps and concepts of the program. A general overview of the objects and entities that play a role in the architecture of \textsc{atomicrex} is provided in \fig{fig:overview}.

Generally, a job can be divided into two parts, the training phase and the output phase. During the training phase selected degrees of freedom (parameters) of the model (usually an interatomic potential) are varied such that the predicted properties (energies, forces, elastic constants, etc., see \sect{sect:properties}) most closely match the target data. 

The training phase is followed by the validation phase. Here, additional properties of interest can be calculated that were not included in the fitting. This allows a convenient separation of  the available data into training and validation sets, where the latter serve to assess the predictive capability of a model.

Once the training process has been completed, the resulting model can be exported in various forms adequate for the use with popular atomistic simulation codes such as \textsc{lammps} \cite{Pli95}.

\section{Key concepts}
\label{sect:key-concepts}

\textsc{atomicrex} operates with two principal types of entities: potentials (models) and structures. 
\begin{itemize}
\item
  A \emph{potential} consists of a parameter set and a specific routine that calculates the total energy and the forces given an atomic structure.
\item
  A \emph{structure} consists of a simulation cell with or without periodic boundary conditions and a list of atoms.
\end{itemize}  
Potentials and structures each possess a set of \emph{degrees of freedom} and a set of \emph{properties}, which will be discussed next.

\subsection{Degrees of freedom}

A degree of freedom (DOF) describes an aspect of a model or structure
that can be continuously varied. The DOFs of a potential are varied
during the training process to minimize the objective function
and optimize the model with respect to the desired properties whereas
the DOFs of an atomic structure may be varied during
structure relaxation, i.e. commonly in order to minimize the potential
energy of a configuration. The calculation of properties for a given set of model parameters may involve relaxation of atomic structures; DOFs of an atomic structure are therefore varied in an inner optimization loop. The DOFs of the model are varied in an outer optimization loop as part of the fitting process.

There exist different classes of DOFs: In the most simple case,
a DOF is a scalar variable that describes a single parameter, e.g., the
parameters of a Lennard-Jones potential or the lattice parameter of a cubic lattice. Furthermore, multi-dimensional DOFs exist that control more complex aspects, e.g., the atomic positions of a structure or the coefficients of a spline potential function.

Each type of potential or structure in \textsc{atomicrex} exposes a certain set of such DOFs. The user has to specify an initial value in the input file for each DOF. While by default every DOF is static, i.e., its value does not change during a job, the user can mark selected DOFs to be included in the fitting process. 

\subsection{Properties}
\label{sect:properties}

Properties are quantities that can be calculated from the DOFs of a structure or model, i.e., they are functionals of the DOFs. The calculation of properties of a structure, e.g., its elastic constants, can be rather complex and involve the evaluation of energy and forces --- possibly multiple times and including an inner relaxation loop. By contrast, properties of potentials can be directly derived from the current values of the DOFs of the model. This complexity is handled efficiently and transparently by the framework of \textsc{atomicrex}.

Each type of structure and potential is associated with a specific set of properties. By default only very few of them are calculated, such as the potential energy and the volume of a structure. However, the user can enable the calculation of additional properties (e.g., elastic constants,  energy differences, lattice parameters, etc.) independently for the
training and/or the final validation stage. When including a property in the training phase, one has to specify a target value for that property. The target value is a scalar in the case of simple properties, e.g., the energy of a structure, or it can be multi-dimensional (vector) data such as the forces acting on the atoms of a structure. As discussed below, each property included in the training set contributes according to its weight to the objective function, which is minimized during the fitting process.

Note that properties are calculated after the DOFs of a structure have
been relaxed. This implies that, if relaxation is enabled for certain
DOFs, the fitting process consists of two nested minimization loops:
Each time the objective function for the potential fitting is
evaluated in the outer optimization loop, the energy of all structures
is first minimized with respect to the structural DOFs. A third level
of variation may exist if the relaxation of the atomic positions has
been enabled. In that case, the atomic positions must be relaxed each
time the total energy is calculated for a given trial configuration of
the DOFs for which relaxation is enabled.

\subsection{Objective function}
\label{overview:objective-function}

The objective (or cost) function is the main quantity being computed
by \textsc{atomicrex} in order to optimize the model toward the target values. 
It can be written in the general form
\begin{align}
\label{eq:objective-function}
f
&=
\underbrace{
\sum_G
\bar{w}_G
\underbrace{
\sum_S
\bar{w}_{GS}
\underbrace{
\sum_P
\bar{w}_{GSP}
r_{GSP}
}_\text{properties}
}_\text{structures}
}_\text{structure groups}
\end{align}
with the \emph{relative weight factors}
\begin{align}
\label{eq:weights}
\bar{w}_{GSP} &= w_{GSP} \Big/ \sum_{P'}^\text{properties}       w_{GSP'} \\
\bar{w}_{GS}  &= w_{GS}  \Big/ \sum_{S'}^\text{structures}       w_{GS'} \\
\bar{w}_{G}   &= w_G     \Big/ \sum_{G'}^\text{structure groups} w_{G'}
,
\end{align}
which can be assigned by the user at the level of structure groups, 
structures, and individual properties providing fine-grained control 
over the importance of different targets during the fitting process.

The residual $r_{GSP}$ is calculated for each property from the predicted, $A^{\text{predicted}}$, and the target value, $A^{\text{target}}$, according to the selected \emph{residual style}, which can be specified by the user. For example, the \verb|squared| residual style implies
\begin{align}
  r_{GSP} &= \left[ \left(
    A^{\text{predicted}}_i - A^{\text{target}}_i\right)
    \big/ \delta_{GSP} \right]^2.
  \label{eq:residual}
\end{align}
Here, $\delta_{GSP}$ is the \emph{tolerance} that has been specified
for the property as explained below.

The rationale behind this nested weight model described by \eq{eq:objective-function} is as follows. When fitting potentials, certain structures are usually more important than others. For example, in the case of silicon the diamond structure should naturally be given a higher weight than, say, the face-centered cubic structure, whereas the opposite applies to e.g., aluminum or gold. To achieve this balance one can set the weights of individual structures ($w_{GS}$) as well as structure groups ($w_G$). Note that these weights are normalized on each individual hierarchy level such that the ratio between structures remains the same when the weights of individual properties are changed.

Similarly, it is advantageous to use ``intuitive'' weights to express
for example the notion that the cohesive energy of a certain structure
is ``three times more important'' than the bulk modulus. These 
properties, however, have very different units and thus the
differences between predicted and target values, $A^{\text{predicted}}
- A^{\text{target}}$, can be of very different magnitudes. It is
usually inconvenient to adjust the property weights manually to
correct for this discrepancy. One can partially remedy the situation by
normalizing the residual by the target value of
the property, essentially making it unitless:
\begin{align}
  r_{GSP} &= \left[ \left(
    A^{\text{predicted}} - A^{\text{target}}\right)
    \big/ A^{\text{target}} \right]^2.
\end{align}
However, this approach obviously has problems when the target value
is zero.

A more refined approach is to use the \emph{tolerance} parameter
$\delta$, introduced in \eq{eq:residual}, that enables one to specify the acceptable range of deviation for each
property. It naturally carries the same unit as the property such that
all residuals are unitless. This allows one to use this parameter
rather sensibly. For example, it is often reasonable to aim for a
cohesive energy to agree within, say, $\delta = 0.1\,\text{eV}$ with
the target value, whereas for an elastic constant $\delta =
5\,\text{GPa}$ could be an acceptable deviation. For convenience \textsc{atomicrex} provides several presets, which assign default tolerance factors based on the unit of each property.

section{Optimization algorithms}
\label{sect:optimization-algorithms}

At present \textsc{atomicrex} \emph{directly} supports the following
local and global optimization algorithms, which are selected in  
the input file:
\begin{itemize}
\item
  the limited-memory Broyden-Fletcher-Goldfarb-Shanno (L-BFGS-B)
  minimizer, which is a popular quasi-Newton method with support for
  bound constraints,

\item
  the ``simply poking around'' (SPA) minimizer, which randomly samples
  the parameter space, and

\item
  all algorithms provided by the \textsc{NLopt} library \cite{nlopt},
  which includes methods for global optimization, local
  derivative-free optimization, and local gradient-based methods.
\end{itemize}

In addition an even larger number of optimization algorithms are accessible via the Python interface (\sect{sect:python-interface}) in combination with third-party libraries such as \textsc{scipy} \cite{scipy} and \textsc{scikit-learn} \cite{scitkit-learn}. In other words, \textsc{atomicrex} provides an efficient compute engine that can be used as a kernel in optimization algorithms implemented in the Python language.

\section{Potentials}
\label{sect:potentials}

\textsc{atomicrex} supports a number of different potential types that at present include e.g,
\begin{itemize}
\item
  the embedded atom method (EAM) \cite{DawBas84},
\item
  the modified embedded atom method (MEAM) \cite{Bas87},
\item
  Tersoff-Abell style potentials \cite{Abe85, Ter86},
\item
  analytic bond-order potentials in the generalized Brenner format
  \cite{Bre90, AlbNorAve02, ErhAlb05, JusErhTra05, ErhJusGoy06},
\item
  the concentration dependent embedded-atom method (CD-EAM) format \cite{CarCroCar05, StuSadErh09, SadErhStu09},
\item
  the angular dependent potential (ADP) format \cite{MisMehPap05}, and
\item
  Stillinger-Weber style potentials \cite{StiWeb85},
\end{itemize}
where the latter two are examples for potential formats that can be generated using the math parser functionality described below.
These types can be combined to develop potentials for complex multi-component systems that cannot be described by a single model.

Within the spectrum of potential types listed above, \textsc{atomicrex} enables the user to choose functional forms with considerable freedom. This functionality is implemented by integrating the \textsc{muparser} math parsing library \cite{muparser}. For example, a general, single-element EAM potential consists of three functionals, each of which can be specified freely by the user in terms of the  mathematical form and the number of parameters (degrees of freedoms) it contains. For multi-element potentials, functionals can be defined once and shared between several (pair-wise or other) interactions. And it is possible to link parameters that occur in more than one functional in order to reduce the effective number of degrees of freedom to be fitted. The user guide contains a number of examples that demonstrate this functionality in practice and exhibit its flexibility.

\section{Structures}
\label{sect:structures}

Structures are one of the principal entities in
\textsc{atomicrex}. They are specified in the main input file and can be 
compiled into groups, which is particularly convenient when dealing with large sets of structures. Each structure exhibits a set of computable properties that can be included in the cost function during training or solely in the validation phase, e.g., in order to assess the predictive power of a model.

For structure specification, the user can resort to a large database of
predefined structures, which includes practically all important crystalline as well as some non-periodic structures for single and two component systems as well as a number of ternary structures. These procedural structures are all parametrized in terms of lattice parameter, axial ratio, internal degrees of freedom, etc. This enables to efficiently relax structures with respect to these structural parameters, e.g. for directly fitting to properties such as the cohesive energy or both clamped and relaxed-ion elastic constants.

In addition, it is possible to include custom structures, which can be either specified directly in the XML input file or imported from external files in standard data formats used by DFT and MD simulation codes. Structure databases can be kept in separate files to reference and re-use them from more than one fitting job.

\section{Derived properties}
\label{sect:derived-properties}

When training potential models it is often key to include information that pertains to (energy) \emph{differences} between structures. This concerns for example point defects (vacancies, interstitials, substitutional atoms), line defects (dislocations), and planar defects (e.g., surfaces, interfaces, antiphase boundaries, or stacking faults) but also structural energy differences (e.g., hexagonal close-packed vs. face-centered cubic or face-centered cubic vs. body-centered cubic).

To this end, \textsc{atomicrex} provides a mechanism that enables the user to define a defective structure (including e.g., a vacancy or an interstitial atom) based on a reference structure, typically represented by a primitive unit cell. The computation of defect formation or binding energies is then accomplished by defining a so-called \emph{derived property}, which links several other properties, often from more than one structure. For example, it is possible for the user to define a defect formation energy by subtracting the reference energy, computed from an ideal unit cell, from the total energy of a supercell containing the defect. \textsc{atomicrex} takes care of computing all input properties that enter into the user-defined formula for the \emph{derived property}, making it possible to readily fit a model to defect formation energies, energy differences between phases, surface energies, or specific phonon frequencies.

\section{Python interface}
\label{sect:python-interface}

\textsc{atomicrex} can be used as a standalone tool and, via its Python interface, also from scripts and other programs. In standalone mode, the program simply processes the XML input file, which completely specifies all inputs to the fitting problem, the optimization method to use, and the output to generate.

The Python interface has been developed to give additional flexibility for customized model development setups and to provide a direct interface with, for example, first-principles codes via the Atomic Simulation Environment (\textsc{ASE}) \cite{BahJac02}. To this end, the Python interface provides full access to the current model parameters, atomic structures, properties, and weights. \textsc{atomicrex} natively supports the \textsc{ASE} atomic structure format for exchange with other frameworks and simulation codes.

Furthermore, the code exposes the current model parametrization
as a linear state vector and allows to evaluate the objective function
in order to enable the rapid implementation of new optimization schemes
and to leverage the wide spectrum of powerful optimization algorithms
that are readily available in the Python ecosystem. 

\section{Conclusions and outlook}
\label{sect:conclusions-and-outlook}

In this paper we have described the \textsc{atomicrex} code, which provides a flexible, extensible, and efficient framework for the construction of atomic scale models suitable for e.g., molecular dynamics and Monte Carlo simulations. \textsc{atomicrex} supports a variety of interatomic potential types, and their functional form can be freely determined by the user via a built-in math parser. The code has been optimized for computational efficiency, enabling larger training and validation sets. While it already includes an extensive database of predefined structures and properties, it also allows the inclusion of custom structures and the definition of complex ``derived'' properties that are based on the combination of several individual properties and/or structures. Finally, \textsc{atomicrex} provides an interface to the Python scripting environment for integration with many advanced scientific libraries available in the Python ecosystem. The code is provided under an open-source license and available via \url{http://atomicrex.org}. We also provide an extensive user guide with many examples and a comprehensive description of features.

\textsc{atomicrex} continues to be developed with an emphasis on code extensibility and speed. In fact, while it already provides an excellent platform not only for the development of potentials using ``classic'' functional forms (EAM, ABOP, MEAM etc.), it can be extended to include e.g., artificial neural network (ANN) potentials \cite{BehPar07, Beh16, ArtUrb16}, tight binding models \cite{LenKreKwo97}, or Gaussian approximation potentials \cite{BarPayKon10}. In this context, we provide a full documentation of the application programming interface (available as part of the Git repository) to enable other researchers to contribute to the development e.g., via new models (potentials) or optimization schemes.

As pointed out above, the Python interface allows easy and seamless integration with various existing libraries for scientific computing and machine learning like \textsc{scipy} \cite{scipy}, \textsc{scikit-learn} \cite{scitkit-learn}, or \textsc{TensorFlow} \cite{tensor-flow}. This opens up the possibility to employ variable training and validation sets for e.g., Bayesian error estimation (see e.g., \cite{FreJacBro04}), or manipulate the parameter vector using genetic algorithms. Finally, via its \textsc{ase} interface, \textsc{atomicrex} can be readily integrated in a dynamic workflow that spans from the generation of reference data using first-principles codes via training and validation of an atomistic model all the way to deployment of the model in molecular dynamics or Monte Carlo simulations. Hence, it can be employed in, for example, on-the-fly scale-bridging simulations \cite{CsaAlbPay04}.

\section*{Acknowledgments}

Funding by the Knut and Alice Wallenberg Foundation, the Deutsche Forschungsgemeinschaft (DFG) through Grant No. STU 611/1-1, the Swedish Research Council, and the Helmholtz Joint Research Group 411 (ODS-HiTs) is gratefully acknowledged. 

\section*{References}

\providecommand{\newblock}{}


\begin{thebibliography}{10}
\expandafter\ifx\csname url\endcsname\relax
  \def\url#1{{\tt #1}}\fi
\expandafter\ifx\csname urlprefix\endcsname\relax\def\urlprefix{URL }\fi
\providecommand{\eprint}[2][]{\url{#2}}

\bibitem{Jon15}
Jones R~O 2015 {\em Reviews of Modern Physics\/} {\bf 87} 897--923

\bibitem{BroGah07}
Brommer P and G\"ahler F 2007 {\em Modelling and Simulation in Materials
  Science and Engineering\/} {\bf 15} 295

\bibitem{JaramilloBotero2014}
Jaramillo-Botero A, Naserifar S and Goddard W~A 2014 {\em Journal of Chemical
  Theory and Computation\/} {\bf 10} 1426--1439

\bibitem{Duff2015}
Duff A~I, Finnis M, Maugis P, Thijsse B~J and Sluiter M~H 2015 {\em Computer
  Physics Communications\/} {\bf 196} 439 -- 445

\bibitem{WarAgrFlo12}
Ward L, Agrawal A, Flores K~M and Windl W 2012 Rapid production of accurate
  embedded-atom method potentials for metal alloys
  \urlprefix\url{https://atomistics.osu.edu/eam-potential-generator/}

\bibitem{ArtUrb16}
Artrith N and Urban A 2016 {\em Computational Materials Science\/} {\bf 114}
  135--150

\bibitem{Pli95}
Plimpton S 1995 {\em J. Comput. Phys.\/} {\bf 117} 1

\bibitem{atomistica}
Atomistica \urlprefix\url{http://www.atomistica.org/}

\bibitem{kim}
Tadmor E~B, Elliott R~S, Sethna J~P, Miller R~E and Becker C~A 2011
  Knowledgebase of interatomic models (kim) \urlprefix\url{https://openkim.org}

\bibitem{TadEllSet11}
Tadmor E~B, Elliott R~S, Sethna J~P, Miller R~E and Becker C~A 2011 {\em JOM\/}
  {\bf 63} 17

\bibitem{nlopt}
\textsc{NLopt}: a free/open-source library for nonlinear optimization
  \urlprefix\url{http://ab-initio.mit.edu/wiki/index.php/NLopt}

\bibitem{scipy}
\textsc{SciPy}: a {P}ython-based ecosystem of open-source software for
  mathematics, science, and engineering \urlprefix\url{https://www.scipy.org/}

\bibitem{scitkit-learn}
\textsc{scikit-learn}: Machine learning in {P}ython
  \urlprefix\url{http://scikit-learn.org/}

\bibitem{DawBas84}
Daw M~S and Baskes M~I 1984 {\em Physical Review B\/} {\bf 29} 6443

\bibitem{Bas87}
Baskes M~I 1987 {\em Physical Review Letters\/} {\bf 59} 2666

\bibitem{Abe85}
Abell G~C 1985 {\em Physical Review B\/} {\bf 31} 6184--6195

\bibitem{Ter86}
Tersoff J 1986 {\em Physical Review Letters\/} {\bf 56} 632--635

\bibitem{Bre90}
Brenner D~W 1990 {\em Physical Review B\/} {\bf 42} 9458--9471

\bibitem{AlbNorAve02}
Albe K, Nordlund K and Averback R~S 2002 {\em Physical Review B\/} {\bf 65}
  195124

\bibitem{ErhAlb05}
Erhart P and Albe K 2005 {\em Physical Review B\/} {\bf 71} 035211

\bibitem{JusErhTra05}
Juslin N, Erhart P, Tr\"askelin P, Nord J, Henriksson K, Salonen E, Nordlund K
  and Albe K 2005 {\em Journal of Applied Physics\/} {\bf 98} 123520

\bibitem{ErhJusGoy06}
Erhart P, Juslin N, Goy O, Nordlund K, M\"uller R and Albe K 2006 {\em Journal
  of Physics: Condensed Matter\/} {\bf 18} 6585--6605

\bibitem{CarCroCar05}
Caro A, Crowson D~A and Caro M 2005 {\em Physical Review Letters\/} {\bf 95}
  075702

\bibitem{StuSadErh09}
Stukowski A, Sadigh B, Erhart P and Caro A 2009 {\em Modelling and Simulation
  in Materials Science and Engineering\/} {\bf 17} 075005

\bibitem{SadErhStu09}
Sadigh B, Erhart P, Stukowski A and Caro A 2009 {\em Philosophical Magazine\/}
  {\bf 89} 3371--3391

\bibitem{MisMehPap05}
Mishin Y, Mehl M and Papaconstantopoulos D 2005 {\em Acta Materialia\/} {\bf
  53} 4029--4041

\bibitem{StiWeb85}
Stillinger F~H and Weber T~A 1985 {\em Physical Review B\/} {\bf 31} 5262--5271

\bibitem{muparser}
\textsc{muparser}: Fast math parser library
  \urlprefix\url{http://muparser.beltoforion.de/}

\bibitem{BahJac02}
Bahn S~R and Jacobsen K~W 2002 {\em Computing in Science and Engineering\/}
  {\bf 4} 56--66

\bibitem{BehPar07}
Behler J and Parrinello M 2007 {\em Physical Review Letters\/} {\bf 98} 146401

\bibitem{Beh16}
Behler J 2016 {\em Journal of Chemical Physics\/} {\bf 145} 170901

\bibitem{LenKreKwo97}
Lenosky T~J, Kress J~D, Kwon I, Voter A~F, Edwards B, Yang D~F, Yang S and
  Adams J~B 1997 {\em Physical Review B\/} {\bf 55} 1528--1544

\bibitem{BarPayKon10}
Bart{\'o}k A~P, Payne M~C, Kondor R and Cs{\'a}nyi G 2010 {\em Physical Review
  Letters\/} {\bf 104} 136403

\bibitem{tensor-flow}
\textsc{TensorFlow}: An open source software library for machine intelligence
  \urlprefix\url{https://www.tensorflow.org/}

\bibitem{FreJacBro04}
Frederiksen S~L, Jacobsen K~W, Brown K~S and Sethna J~P 2004 {\em Physical
  Review Letters\/} {\bf 93} 165501

\bibitem{CsaAlbPay04}
Csanyi G, Albaret T, Payne M~C and De~Vita A 2004 {\em Physical Review
  Letters\/} {\bf 93} 175503

\end{thebibliography}
\end{document}